\documentclass[letterpaper]{ptephy} 

\usepackage{amsmath} 
\usepackage{boites}

\usepackage{graphicx} 

\begin{document}

\title{Quantum Twist to Complementarity: A Duality Relation}
\author{Tabish Qureshi\footnote{E-mail: tabish@ctp-jamia.res.in}}
\address{Centre for Theoretical Physics, Jamia Millia Islamia,
New Delhi, India.}

\begin{abstract}
Some recent works have introduced a quantum twist to the concept of
 complementarity, exemplified by a setup in which the which-way detector is
in a superposition of being present and absent. It has been argued that such
experiments allow measurement of particle-like and wave-like behavior at
the same time. Here we derive an inequality which puts a bound on the
visibility of interference and the amount of which-way information that one
can obtain, in the context of such modified experiments. As the wave-aspect
can only be revealed by an ensemble of detections, we argue that in such
experiments, a single detection can contribute only to one subensemble,
corresponding to either wave-aspect or particle aspect. This way, each
detected particle behaves either as particle or as wave, never both,
and Bohr's complementarity is fully respected.
\end{abstract}


\maketitle

\section{Introduction}

The two-slit experiment carried out with particles is a testbed of various
foundational ideas in quantum theory. It has been used to exemplify
wave-particle duality and Bohr's complementarity principle \cite{bohr}.
The two-slit experiment captures the essence of quantum theory in such
a fundamental way that Feynman went to the extent of stating that it is a
phenomenon ``which has in it the heart of quantum mechanics; in reality it
contains the {\em only} mystery" of the theory \cite{feynman}.

\begin{figure}[h!]
\centering
\includegraphics[width=4.0 in]{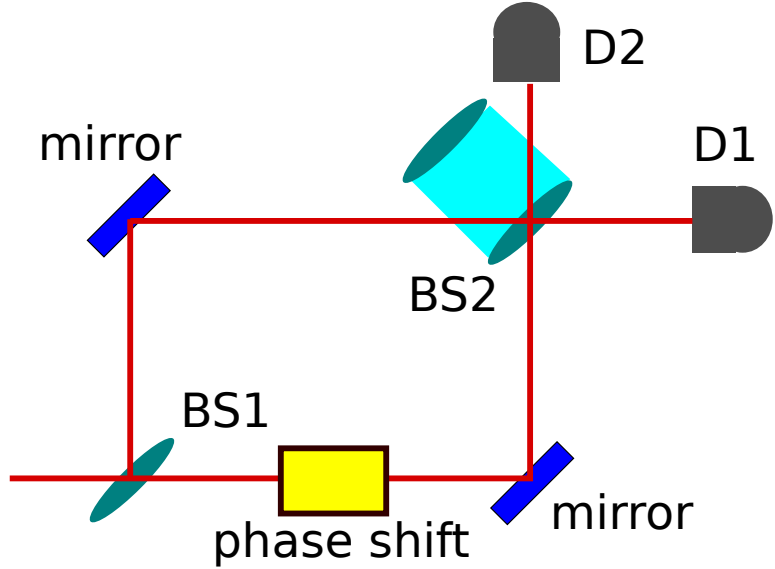}
\caption{Schematic diagram to illustrate a typical experiment to test
complementarity with an introduction of a quantum device BS2. The beam-splitter
BS2 is in a superposition of being present in the path of the photon and
being away from it.}
\label{qcsetup}
\end{figure}

Neils Bohr had stressed that the wave-nature of particles, characterized
by interference, and the particle-nature, characterized by the knowledge
of which slit the particle passed through, are mutually exculsive. He
argued that in a single experiment, one could see only one of these two 
complementary properties at a time. Niels Bohr elevated this concept to the
status of a separate principle, the principle of complementarity \cite{bohr}.
Two-slit interference experiment with a which-way detector and
the Mach-Zehnder interferometer have been extensively used to study 
complementarity. It has been demonstrated in many different kinds of 
experiments, that if one obtains the which-path information about a
particle, the interference pattern cannot be obtained. Conversely, if
the interference pattern is obtained, the which-path information is
necessarily destroyed. In some clever experiments, the experimenter can
choose to obtain the which-path information, much after the particle has
been registered on the screen \cite{kim,jacques}. Such ``delayed-choice"
experiments present several conceptual difficulties. Nevertheless, it
is generally accepted that in an experiment, only one of the two aspects
can be seen at a time, and that the two are mutually exclusive.

\section{Quantum twist to complementarity}

Recently a new kind of experiments to test complementarity have been
proposed \cite{terno} and carried out \cite{celeri,roy,tang,peruzzo,kaiser}, where the which-way
detector is a quantum device which is prepared in a superposition of 
being present and absent. A typical such experiment is shown in FIG.
\ref{qcsetup}.
Here, if the beamsplitter BS2 is absent, the two photon detectors will
give which-way information about every photon detected. The BS2 is
present, the two paths are mixed and which-way information is lost.
However, the phases of the two paths can be tuned in such a way that
they destructively interfere at (say) detector $D1$. Detector $D1$
not detecting any photons indiciates interference. Now the setup
is modified in such a way that BS2 is in a superposition of two locations,
one of which is in the path of the photon, and the other is outside it.

If $|N\rangle$ represents the state of BS2 when it is in the path of the
photon, and $|Y\rangle$ represents the state when it is outside it, the
argument is that the combined state of BS2 and the photon can be written as
\cite{terno,celeri}
\begin{equation}
|\psi\rangle = \sqrt{c}|Y\rangle|\text{particle}\rangle_S
          + \sqrt{1-c}|N\rangle|\text{wave}\rangle_S ,
\end{equation}
where $|\text{particle}\rangle_S$ represents the state of the photon where
it behaves like a particle and $|\text{wave}\rangle_S$ represents its state 
when it behaves like a wave. The claim is that the wave and particle nature of
the photon is present at the same time, in a superposition, which allows
one to get some more information as compared to the conventional which-way
experiments \cite{terno,celeri}.

Here we carry out a detailed analysis of a {\em gedanken} setup which
represents a typical such experiment, to
explore what information such experiments can yield.

\section{Path-distinguishability and fringe visibility}

\begin{figure}[h!]
\centering
\includegraphics[width=4.5 in]{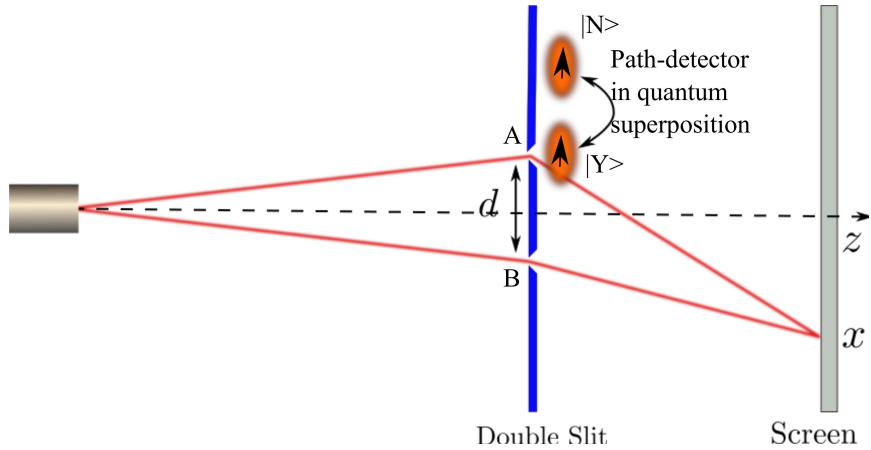}
\caption{A two-slit experiment with a one-bit path-detector in front of slit A.
The one-bit detector is in a superposition of being present in the path of
the photon and being away from it.}
\label{q2slit}
\end{figure}

Consider a conventional two slit experiment with particles, with a single-bit
which-way detector sitting in the path of slit A (see FIG. \ref{q2slit}).
The which-way detector is
initially in the state $|d_2\rangle$. If the particle passes through
slit A, the which-way detector comes to a state $|d_1\rangle$. Corresponding
to the particle passing through slit B, the detector remains in the
state $|d_2\rangle$.  We can define the distinguishability of the two paths by
${\mathcal D} = (1 - |\langle d_1|d_2\rangle|)$,
where $|d_1\rangle$ and $|d_2\rangle$ are assumed to be normalized,
but not necessarily orthogonal to each other.
Clearly, for completely orthogonal $|d_1\rangle$ and $|d_2\rangle$,
${\mathcal D} = 1$, and for identical $|d_1\rangle$ and $|d_2\rangle$,
${\mathcal D} = 0$. If $|d_1\rangle$ and $|d_2\rangle$ are orthogonal
to each other, one can find an observable of the detector, for which
the two states can give two distinct eigenvalues. Measuring such an
observable, one can find out which of the two slits the particle
went through. 

Next we allow our which-way detector to be a quantum object in the sense
that it can be in a superposition of two locations. The state $|Y\rangle_L$
corresponds to the detector being in front of slit A, and $|N\rangle_L$
corresponds to it being away from slit A. Let the detector be prepared in
an intial state
\begin{equation}
|\phi_0\rangle_D = |d_2\rangle\left(\sqrt{c}|Y\rangle_L
                   + \sqrt{1-c}|N\rangle_L\right) ,
\label{phi0}
\end{equation}
where $c$ is a real constant between 0 and 1. The value of $c$ being 1
means that the which-way detector is in front of slit A, and being 0
means it is away from the slit. The states $|d_1\rangle,~|d_2\rangle$
can give which-way information only when the which-way detector is
in front of slit A. Keeping this in mind, we define the which-way 
distinguishability as
\begin{equation}
{\mathcal D} = (1 - |\langle d_1|d_2\rangle|)~
{_D\langle\phi_0|(|Y\rangle\langle Y|)_L|\phi_0\rangle_D}.
\end{equation}
Here we have assumed that when a particle passes through the double slit,
the path through slit A gets correlated to $|d_1\rangle$, and that through
slit B get correlated to $|d_2\rangle$.
For the state given by (\ref{phi0}), distinguishability takes the form
\begin{equation}
{\mathcal D} = c(1 - |\langle d_1|d_2\rangle|).
\label{D}
\end{equation}
As one can
see, the two paths will be fully distinguishable when $c=1$ and 
$\langle d_1|d_2\rangle=0$.

Let us now assume that a particle traveling along the z-direction passes
through the double-slit, with a slit separation $d$, and also interacts with
a which-path detector.
Through a unitary process, the
detector states get correlated with the states of the particle coming
out of the two slits. The combined state of the particle and the which-path
detector, when the particle emerges from the double-slit (time $t=0$),
is assumed to have the form
\begin{eqnarray}
\Psi(x,0) &=& A\sqrt{c} \left(|d_1\rangle e^{-{(x-d/2)^2\over 4\epsilon^2}}
+ |d_2\rangle e^{-{(x+d/2)^2\over 4\epsilon^2}}\right)|Y\rangle_L\nonumber\\
&&+ A\sqrt{1-c} \left(e^{-{(x-d/2)^2\over 4\epsilon^2}}
+ e^{-{(x+d/2)^2\over 4\epsilon^2}}\right)|d_2\rangle |N\rangle_L
\label{entstate}
\end{eqnarray}
where $A = {1\over\sqrt{2}}(2\pi\epsilon^2)^{-1/4}$. Here we assume the
states of the particle coming out of the slits to have a Gaussian form,
with a width $\epsilon$, centered at $\pm d/2$.
We do not expicitly consider the dynamics of the particle in
the z-direction. We just assume that the wave-packets are moving in the
positive z-direction with an average momentum $p_0=h/\lambda_d$, where 
$\lambda_d$ is the d'Broglie wavelength of the particle.
Thus the distance $L$ travelled by the particle in a time $t_L$ is given
by $L = {h\over m\lambda_d}t_L$. This can be rewritten as
$\hbar t_L/m = \lambda_dL/2\pi$.

After a time $t$, the state of the particle and the detector evolves to
\begin{eqnarray}
\Psi(x,t) &=& A_t\sqrt{c} \left(|d_1\rangle e^{-{(x-d/2)^2\over 4\epsilon^2+2i\hbar t/m}}
+ |d_2\rangle e^{-{(x+d/2)^2\over 4\epsilon^2+2i\hbar t/m}}\right)|Y\rangle_L\nonumber\\
&+& A_t\sqrt{1-c} \left(e^{-{(x-d/2)^2\over 4\epsilon^2+2i\hbar t/m}}
+ e^{-{(x+d/2)^2\over 4\epsilon^2+2i\hbar t/m}}\right)|d_2\rangle |N\rangle_L,
\end{eqnarray}
where $A_t = {1\over\sqrt{2}}[\sqrt{2\pi}(\epsilon+i\hbar t/2m\epsilon)]^{-1/2}$.
The probability of finding the particle at position $x$ on the screen is
given by
\begin{eqnarray}
|\Psi(x,t)|^2 &=& |A_t|^2 \left(e^{-{(x-d/2)^2\over 2\sigma_t^2}}
+ e^{-{(x+d/2)^2\over 2\sigma_t^2}}\right.\nonumber\\
&&+(1-c+c\langle d_1|d_2\rangle ) e^{-{x^2+d^2/4\over 2\sigma_t^2}}
e^{{ixd\hbar t/2m\epsilon^2\over 2\sigma_t^2}}\nonumber\\
&&+\left.(1-c+c\langle d_2|d_1\rangle ) e^{-{x^2+d^2/4\over 2\sigma_t^2}}
e^{-{ixd\hbar t/2m\epsilon^2\over 2\sigma_t^2}}\right) ,
\end{eqnarray}
where $\sigma_t^2 = \epsilon^2 + (\hbar t/2m\epsilon)^2$. Writing
$\langle d_2|d_1\rangle$ as $|\langle d_2|d_1\rangle|e^{i\theta}$, and
putting $\hbar t/m = \lambda_dL/2\pi$, the above can be simplified. Further,
for simplicity we put $\theta=0$, which reduces the above to
\begin{eqnarray}
|\Psi(x,t)|^2 &=& 2|A_t|^2 e^{-{x^2+d^2/4\over 2\sigma_t^2}}
\cosh(x d /2\sigma_t^2)\times \nonumber\\
&&\left(1+(1-c+c|\langle d_1|d_2\rangle|)
{\cos\left({{xd\lambda_dL/2\pi \over 4\epsilon^4+(\lambda_dL/2\pi)^2}}\right)\over
\cosh(x d /2\sigma_t^2)} \right)
\label{pattern}
\end{eqnarray}
Eqn.(\ref{pattern}) represents an interference pattern with a fringe width
given by 
\begin{equation}
w = 2\pi\left({(\lambda_dL/2\pi)^2 + 4\epsilon^4\over\lambda_ddL/2\pi}\right)
= {\lambda_dL\over d} + {16\pi^2\epsilon^4\over\lambda_ddL}.
\end{equation}
For $\epsilon^2 \ll \lambda_dL$ we get the familiar Young's double-slit
formula $w \approx \lambda_dL/d$.

Visibility of the interference pattern is conventionally defined as 
\begin{equation}
{\mathcal V} = {I_{max} - I_{min} \over I_{max} + I_{min} } ,
\end{equation}
where $I_{max}$ and $I_{min}$ represent the maximum and minimum intensity
in neighbouring fringes, respectively. In reality, fringe visibility will
depend on many things, including the width of the slits. For example, if
the width of the slits is very large, the fringes may not be visible at all.
Maxima and minima of (\ref{pattern}) will occur at points where the 
value of cosine is 1 and -1, respectively. The visibility can then be 
written down as
\begin{equation}
{\mathcal V} = {1-c+c|\langle d_1|d_2\rangle|\over \cosh(x d /2\sigma_t^2)}.
\end{equation}
Since $\cosh(y) \ge 1$, we get
\begin{equation}
{\mathcal V} \le 1 -c + c|\langle d_1|d_2\rangle|.
\label{visibility}
\end{equation}
Using (\ref{D}) the above equation gives a very important result
\begin{equation}
{\mathcal V} + {\mathcal D} \le 1.
\label{duality}
\end{equation}
Eqn. (\ref{duality}) can be considered as a quantitative statement of
Bohr's complementarity principle. It sets a bound on the which-path 
distinguishability and the visibility of interference that one can
obtain in a single experiment. It is similar in spirit to the well-known
Englert-Greenberger duality relation \cite{greenberger,englert}, but clearly
different from it.

In the context of the quantum twist to which-way experiments,
if $|d_1\rangle$ and $|d_2\rangle$ are orthogonal,
(\ref{visibility}) tells us that the visibility ${\mathcal V}$ can 
at the most be $1-c$. However, in that situation the path distinguishability
${\mathcal D}$, given by (\ref{D}), is $c$. So, even though the which-way
detector is replaced by an equivalent quantum device, it doesn't allow
one to obtain simultaenously the which-path information and the interference,
more precisely than (\ref{duality}). In the special case $c=1$, the relation
(\ref{duality}) will describe the bounds on distinguishability and visibility
in the {\em conventional} complementarity experiments.

If one were to correlate every particle detected on the screen with 
a measurement result on the states $|Y\rangle_L$ and $|N\rangle_L$,
every click will either give a which-way information or it won't. Only those
clicks which do not yield which-way information, will contribute to the
interference pattern. So, which-way information and interference remain
mutually exclusive. Of course, one can obtain a fuzzy which-way information,
but it will result in a fuzzy (not sharp) interference pattern.

\section{A random quantum-eraser}

We now describe a {\em gedanken} experiment based on the so-called quantum
eraser \cite{eraser,eraser1}, which also achieves which-way information
and interference in the same experimental setup, although with some
difference. Let there be a setup of two-slit experiment, with a one-bit
which-way detector in front of one of the slits. The state that comes out
of the slit can be written as
\begin{equation}
\Psi(x) = {1\over\sqrt{2}}(|d_1\rangle \psi_A(x) + |d_2\rangle \psi_B(x)),
\label{entangz}
\end{equation}
where $|d_1\rangle,~|d_2\rangle$ are two orthonormal states of the
which-way detector, and $\psi_A(x),~\psi_B(x)$ are the wave-packets 
emerging from slits A and B, respectively. The probability of a particle
falling at a position $x$ on the screen is given by
\begin{equation}
|\Psi(x)|^2 = {1\over{2}}(|\psi_A(x)|^2 + |\psi_B(x)|^2),
\end{equation}
which gives no interference because of the mutual orthogonality of
$|d_1\rangle,~|d_2\rangle$. Measuring an observable, call it $\sigma_z$,
of the which-way detector whose eigenstates are $|d_1\rangle,|d_2\rangle$,
in coincidence with detected particles, will give which-way information
about each particle. The particle nature is brought out in such measurements.
Let us now imagine that there is another observable,
call it $\sigma_x$, whose eigenstates are $|d_+\rangle,|d_-\rangle$
such that
\begin{equation}
|d_+\rangle = (|d_1\rangle+|d_2\rangle)/2,~~~~
|d_-\rangle = (|d_1\rangle-|d_2\rangle)/2.
\end{equation}
In terms of these states, (\ref{entangz}) can be written as
\begin{equation}
\Psi(x) = {|d_+\rangle\over 2}(\psi_A(x) + \psi_B(x))
        + {|d_-\rangle\over 2}(\psi_A(x) - \psi_B(x)).
\label{entangx}
\end{equation}
Now if one measures the observable $\sigma_x$ in coincidence with the
detected particles, the particles in coincidence with $|d_+\rangle$ will
show an interference pattern, and those in coincidence with $|d_-\rangle$,
will show a $\pi$-shifted interference pattern. The wave nature is brought
out in such measurements.  The measurement on the one-bit which-way detector
can be done {\em after} the particle is detected, and one can choose to
bring out either particle nature or wave nature. The wave nature and the
particle nature are both present at the same time, until the measurement
on the one-bit which-way detector is made.
This is an example of the so-called quantum eraser \cite{eraser,eraser1}.

Now, suppose one randomly measures $\sigma_z$ or $\sigma_x$ in coincidence
with the particles falling on the screen. After detecting all the particles,
one can separate out those which were in coincidence with $\sigma_z$
and $\sigma_x$, respectively. Thus, particle nature and wave-nature can
be explored in the same experiment. The result is quite similar to what 
has been proposed and shown in Refs. \cite{terno,celeri,roy,tang,peruzzo,kaiser}.
The major way in which the 
scheme \cite{terno,celeri,roy,tang,peruzzo,kaiser} is different from our quantum eraser scheme
is that in the former the choice between the particle nature and the
wave nature is randomly made by the quantum nature of the position of
the detector, whereas in the latter it is made by the experimenter.
In both the schemes, every single particle detected has to clearly follow
either wave nature or particle nature. Another difference is that in the
quantum eraser scheme the which-way information is always carried by the
which-way detector, but can be ``erased" by the choice of the observable
of the which-way detector which is measured. In the scheme of
\cite{terno,celeri,roy,tang,peruzzo,kaiser}, the which-way information is there only
with a probability (say) $c$; there is a probability $1-c$ that there is
no which-way information.

\section{Conclusions}

In conclusion, we have theoretically analyzed the quantum twist to
complementarity introduced recently in the context of some modified
interference experiments. We have derived a duality relation in the context
of such experiments. It puts a bound on the which-way information that
one can extract, and the visibility of interference in the same experiment.
We emphasize that Bohr's complementarity continues to hold, and should be
viewed in the context of individual outcomes. In each outcome, one can
either get which-way information, in which case particle nature will
emerge, or the detection will contribute to the interference pattern,
which shows wave nature. Although interference builds up after registering
many particles, only those detections will contribute to interference
for which no which-way information was found. All those detections
where which-way information is found, will not contribute to the
interference pattern. Thus, which-way information
and interference remain mutually exclusive, although in the new clever
experiments \cite{terno,celeri,roy,tang,peruzzo,kaiser} both aspects can be seen in
a single experimental setup. It is always possible to do imprecise
which-way measurement and get a fuzzy interference for the same particle.
But this was already known before, and has been the motivation
for the Englert-Greeenberger duality relation. Lastly, the new experiments
where a quantum device is added to the interference experiments, do
help us understand Bohr's complementarity principle better.

\ack{
We are thankful to Lucas Chibebe C\'{e}leri for bringing this new aspect
of complementarity to our notice. This work is supported by the University
Grants Commission, India.}

\end{document}